\documentclass[twoside,english,iop]{emulateapj}
\usepackage[T1]{fontenc}
\usepackage{units}
\usepackage{textcomp}
\usepackage{graphicx}

\makeatletter

\providecommand{\tabularnewline}{\\}

\usepackage{apjfonts}
\usepackage[hyperfootnotes=false,colorlinks=true,citecolor=blue]{hyperref}

\shorttitle{STRUCTURE OF $\lambda$ ORI NEBULA}
\shortauthors{LEE ET AL.}

\makeatother

\usepackage{babel}
\begin{document}

\title{IS DUST CLOUD AROUND $\lambda$ ORIONIS A RING OR A SHELL, OR BOTH? }

\author{Dukhang Lee\altaffilmark{1,2}, Kwang-Il Seon\altaffilmark{1,2},
\and Young-Soo Jo\altaffilmark{1}}

\altaffiltext{1}{Korea Astronomy and Space Science Institute, Daejeon 305-348, Republic of Korea; lee.dukhang@gmail.com}
\altaffiltext{2}{University of Science and Technology, Daejeon 305-350, Republic of Korea} 
\begin{abstract}
The dust cloud around $\lambda$ Orionis is observed to be circularly
symmetric with a large angular extent ($\approx$ 8\textdegree ).
However, whether the three-dimensional (3D) structure of the cloud
is shell- or ring-like has not yet been fully resolved. We study the
3D structure using a new approach that combines a 3D Monte Carlo radiative
transfer model for ultraviolet (UV) scattered light and an inverse
Abel transform, which gives a detailed 3D radial density profile from
a two-dimensional column density map of a spherically symmetric cloud.
By comparing the radiative transfer models for a spherical shell cloud
and that for a ring cloud, we find that only the shell model can reproduce
the radial profile of the scattered UV light, observed using the S2/68
UV observation, suggesting a dust shell structure. However, the inverse
Abel transform applied to the column density data from the Pan-STARRS1
dust reddening map results in negative values at a certain radius
range of the density profile, indicating the existence of additional,
non-spherical clouds near the nebular boundary. The additional cloud
component is assumed to be of toroidal ring shape; we subtracted from
the column density to obtain a positive, radial density profile using
the inverse Abel transform. The resulting density structure, composed
of a toroidal ring and a spherical shell, is also found to give a
good fit to the UV scattered light profile. We therefore conclude
that the cloud around $\lambda$ Ori is composed of both ring and
shell structures.
\end{abstract}


\section{INTRODUCTION}

It has been reported that the O8 III star $\lambda$ Orionis, a member
of the Orion OB1 association (\citealt{Murdin1977}), excites a fairly
symmetric \ion{H}{2} region Sh2-264 surrounding it (\citealt{Sharpless1952,Sharpless1959}).
The \ion{H}{2} region is thought to be a good example of a classical
Strömgren sphere, which is spherically symmetric and centered on $\lambda$
Ori (\citealt{Coulson1978,Maddalena1987}). Many observational studies
have found dark clouds external to this \ion{H}{2} region; the clouds
appear to be circularly symmetric, with a large angular extent of
about 8\textdegree{} (\citealp{Wade1957}, \citeyear{Wade1958}; \citealt{Court=0000E8s1972,Coulson1978,Maddalena1987,Malone1987,Zhang1989,Lang2000})
However, there has been a debate as to whether the symmetric appearance
of the cloud is a projection effect of a ``spherical'' shell or
is due to a ``toroidal'' ring. 

\citet{Wade1957} found an expanding, dense shell of neutral hydrogen
surrounding the \ion{H}{2} region. Using the star counts technique,
\citet{Coulson1978} proposed that the dark clouds consist of two
spherical shells: one outside of the \ion{H}{2} region (an ``outer''
shell) and the other within the region (an ``inner'' shell). Their
results, however, were obtained from only northeast quadrant observation
and thus cannot represent the entire structure of the dark cloud system.
\citet{Morgan1980} obtained the optical properties (albedo $a$ and
asymmetric scattering phase function $g$) of the dust clouds by comparing
ultraviolet (UV) observations with results from a multiple scattering
model based on the Monte Carlo radiative transfer method. They assumed
a simple spherical dust shell, a central core, and a diffuse dust
layer filling the space between the shell and the core, and found
plausible results to support the idea that the cloud had a shell structure.
Infrared Astronomical Satellite (IRAS) data on the dust clouds surrounding
$\lambda$ Ori was presented by \citet{Malone1987}. The authors interpreted
the dust clouds as a large expanding shell, which is fragmented into
clumps. For the same IRAS data, \citet{Zhang1989} not only applied
a more refined reduction technique, but also conducted a much more
detailed analysis than was done in the previous study. Despite the
cloud\textquoteright s clear ring-like appearance, they concluded
that a shell structure of the cloud provides a more consistent explanation
for the infrared (IR) emission, especially the diffuse infrared (IR)
features observed within the nebula. 

Unlike the above interpretations, the CO $(J=1\rightarrow0)$ survey
revealed that the dust cloud surrounding $\lambda$ Ori coincides
well with a toroidal ring of molecular clouds (\citealt{Maddalena1987}).
The CO data showed a detailed structure of expanding molecular clouds:
a ring-like structure composed of discrete cloud patches, having little
molecular medium within the ring. In addition to the clear ring molecular
cloud, \citet{Maddalena1987} predicted the existence of an \ion{H}{1}
shell swept up by the expanding \ion{H}{2} region, though this idea
was mostly based on their theoretical models, which had certain assumptions.
Over a decade later, \citet{Lang2000} provided a new CO map that
had a significantly higher sensitivity and much larger area coverage
than those of the previous map. The new survey not only confirmed
the presence of the dense dark globules previously found by \citet{Maddalena1987},
but also revealed diffuse, low mass clouds, some of which extend to
the inside of the ring. In spite of the diffuse medium and, consequently,
the possibility of the existence of a shell, \citet{Lang2000} concluded
that both the IRAS and CO data indicate that the \ion{H}{2} region
is surrounded by a remarkably complete ring. 

The purpose of this paper is to understand the 3D structure of the
dust clouds in the $\lambda$ Ori system. Using a new approach that
combines a 3D Monte Carlo radiative transfer model and the inverse
Abel transform, we investigate whether the $\lambda$ Ori nebula has
a ring or a shell structure, or even both. In Section 2, we describe
the data of the ultraviolet (UV) measurements and the dust reddening.
Section 3 presents the results of a dust radiative transfer simulation
for the shell and the ring models, and the calculation of the inverse
Abel transform. The best-fit model for the structure of the $\lambda$
Ori nebula is given in Section 4. We discuss, in Section 5, an integrated
view of the nebula structure and the applicability of our new methodological
approach to bubbles in the interstellar medium (ISM). A summary is
presented in Section 6.

\section{DATA}

We use measurements of the Orion region obtained with the S2/68 ultraviolet
sky-survey telescope (\citealt{Boksenberg1973}) in four wavebands
(2740, 2350, 1950, and 1550 Å) and compare the data with that obtained
using the dust scattering models. More specifically, ratios of the
observed UV surface brightness to the incident stellar flux, averaged
over 0\textdegree .5 intervals of radius from $\lambda$ Ori, are
utilized (see Table 2 in \citealt{Morgan1980}). The UV surface brightness,
mainly attributed to light scattered by the dust cloud, was measured
after subtracting terrestrial contributions and signals of geocoronal
$L\alpha$ radiation and bright stars. Unfortunately, both the All-sky
Imaging Survey (AIS) and the Medium Imaging Survey (MID) of the GALEX
data (\citealt{Morrissey2007}), as well as the SPEAR/FIMS data (\citealt{Seon2011}),
do not cover the entire target field. Although the S2/68 measurements
were made several decades ago, these measurements were found to be
adequate to investigate the 3D structure of the cloud. 

\begin{figure}[t]
\begin{centering}
\includegraphics[width=8.4cm]{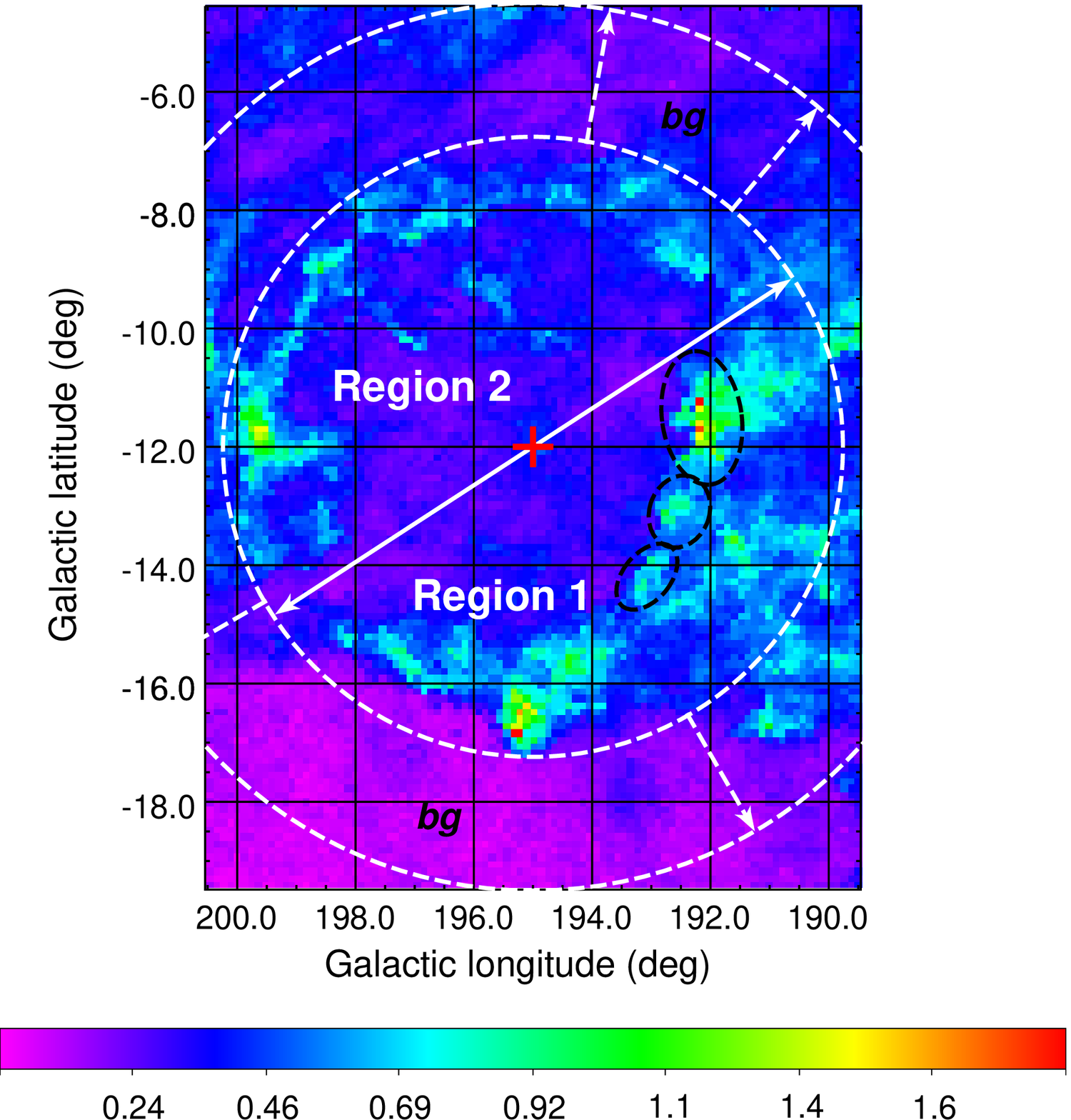}
\par\end{centering}

\protect\caption{\label{Fig1} $E(B-V)$ map of the $\lambda$ Ori nebula region, derived
from Pan-STARRS1 stellar photometry \citep{Schlafly2014a}. $\lambda$
Ori is marked by a red plus sign (+) near $\left(l,b\right)$$ $
= (195\textdegree , $-$12\textdegree ) at the center of the nebula.
According to \citet{Lang2000}, the $\lambda$ Ori nebula can be divided
into two classes: dense and massive globules in Region 1, and diffuse
and low mass clouds in Region 2. The Barnard object B30 is located
at the Galactic coordinates (192\textdegree .3, $-$11\textdegree .4).
Black dashed ellipses are the ``B30 arc complex'' (discussed in
Section 4). The two areas marked as ``$bg$'' indicate background
regions, in which none of $\lambda$ the Ori clouds and little foreground
or background ISM are located. }
\medskip{}
\end{figure}

An extinction map of the $\lambda$ Ori region was obtained from the
Pan-STARRS1 data (\citealt{Schlafly2014a}). This survey map has an
unprecedented resolution compared to that of earlier direct measurements
of dust reddening and covers over three-quarters of the sky including
our target field. The $E(B-V)$ map for the $\lambda$ Ori nebula
region (Figure \ref{Fig1}), assuming a ratio of total to selective
extinction of $R_{V}=A_{V}/E(B-V)=3.1$ (\citealt{Fitzpatrick1999}),
is used to derive a map of optical depth at visible wavelength. A
column density map estimated from the optical depth is converted to
a radial density profile using the inverse Abel transform method (\citealt{Binney2008}).
The technique is useful when the cloud is spherically symmetric and
the dust density depends only on the radial distance from the cloud
center.

\begin{figure*}
\begin{centering}
\includegraphics[width=5.5cm]{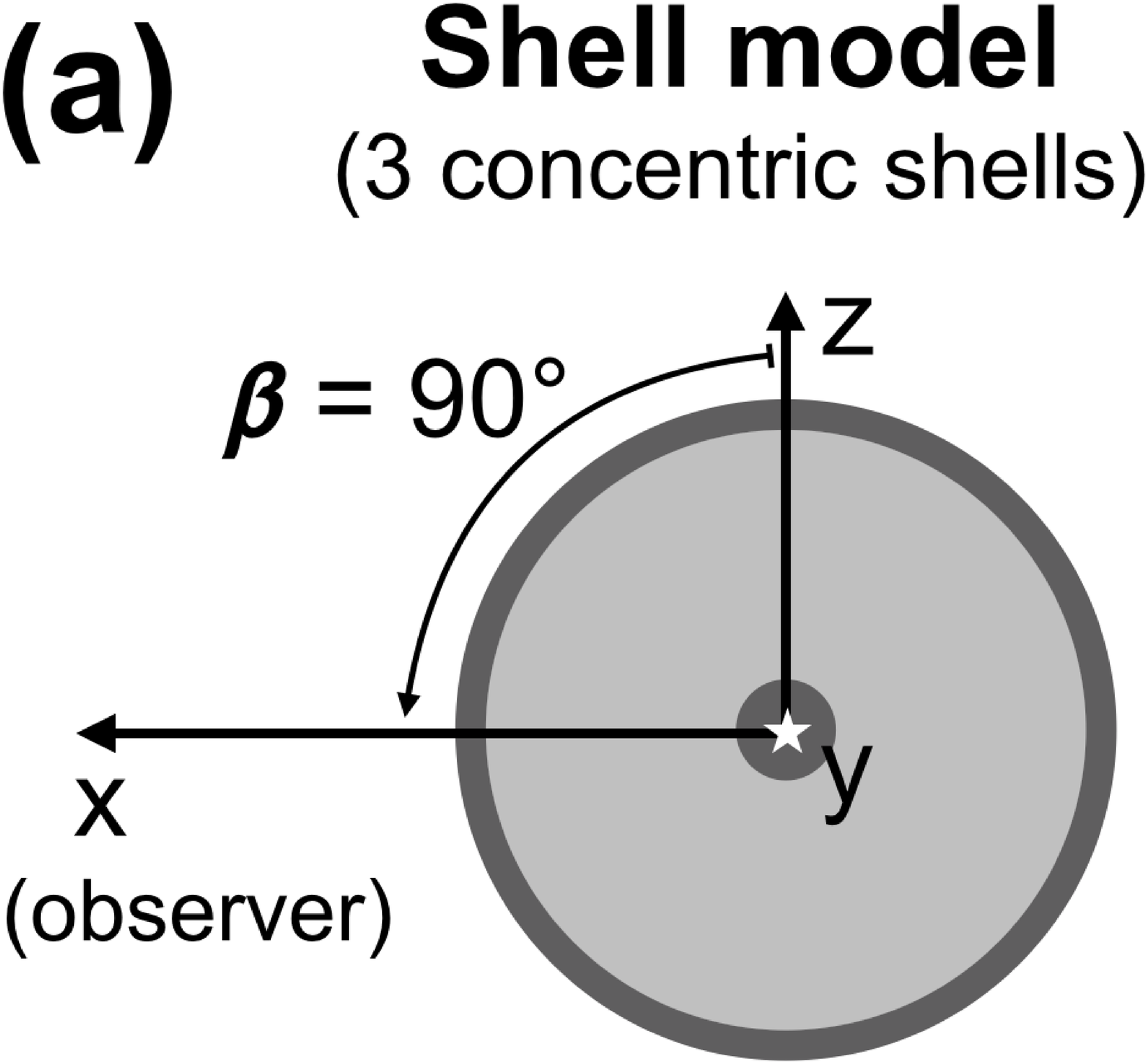}\includegraphics[width=5.5cm]{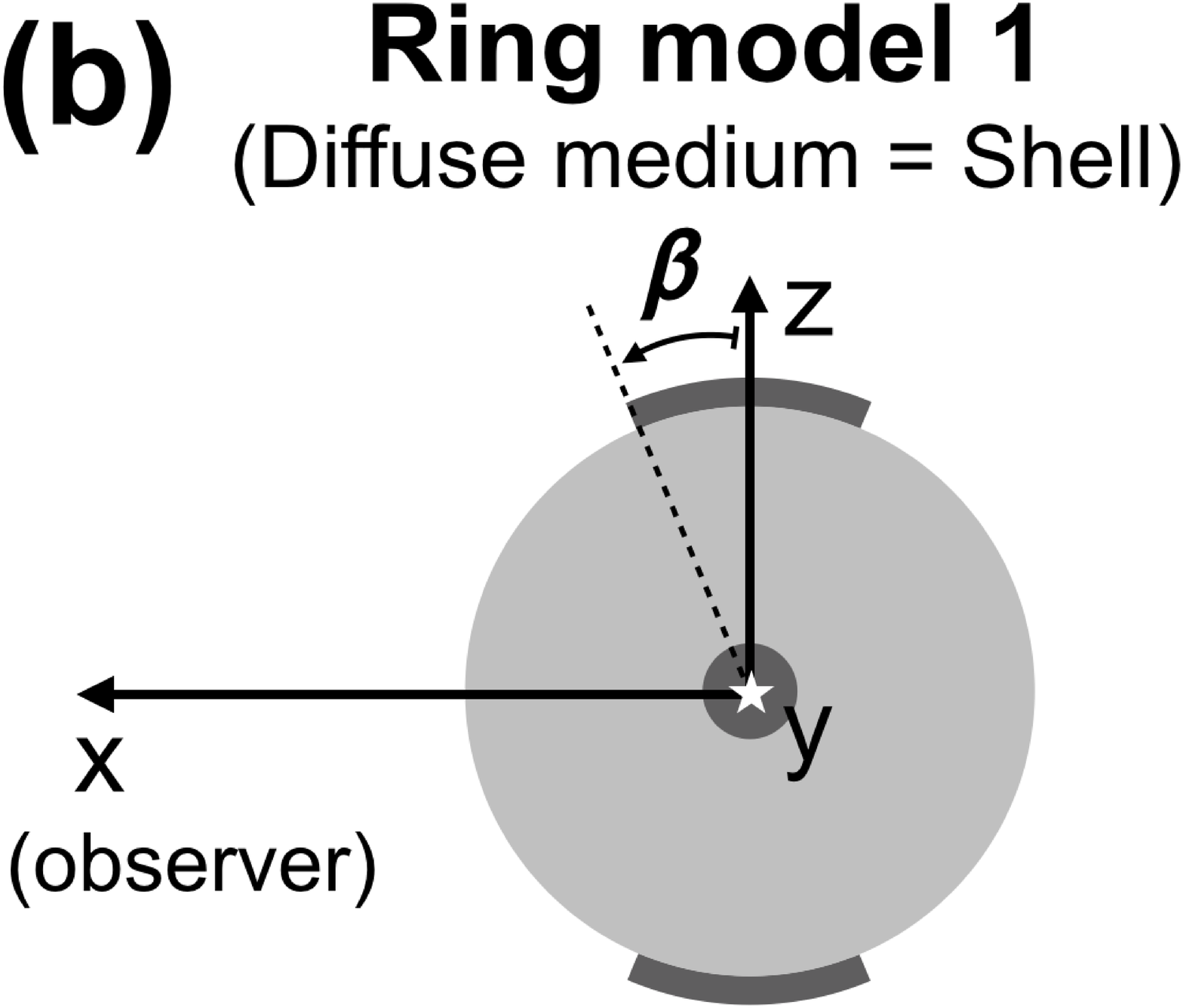}\includegraphics[width=5.5cm]{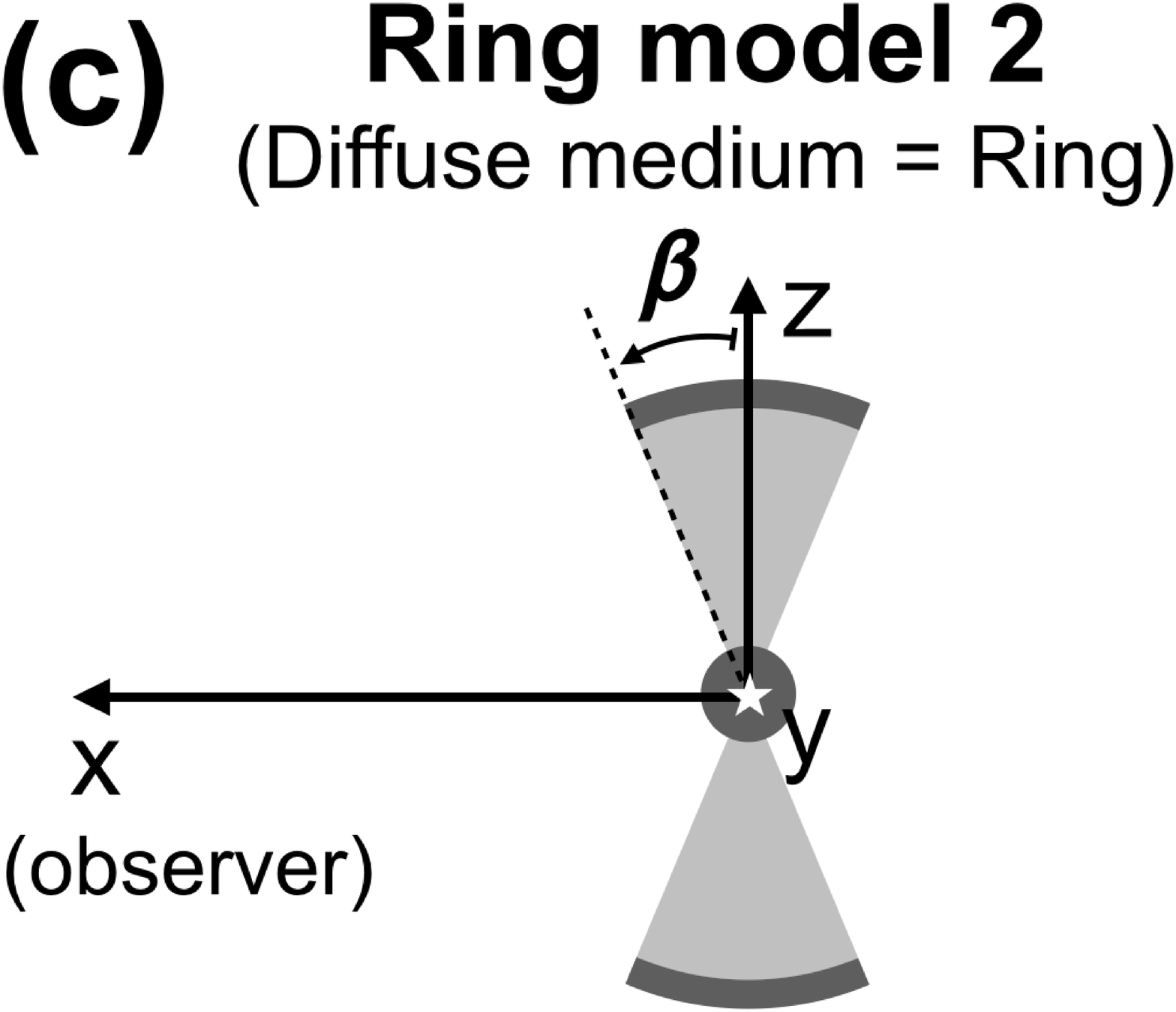}
\par\end{centering}

\medskip{}

\protect\caption{\label{Fig2}Three different dust cloud geometries centered on $\lambda$
Ori: (a) the spherical shell model, (b) ring model 1, which has a
spherically symmetric, low-density medium in addition to a ring-like
dense structure, and (c) ring model 2, in which the low-density medium
has an opening angle $\beta$ < 90\textdegree , together with a ring-like
dense medium as in ring model 1. The three models exhibit rotation
symmetry about the $x$-axis. Each model can be defined by the opening
angle $\beta$. The figures are seen toward the direction of the negative
$y$-axis, on the $x$-$z$ plane, in Cartesian coordinates. }
\medskip{}
\end{figure*}

\section{TEST MODELS FOR SHELL OR RING STRUCTURE}

A \textbf{Mo}nte \textbf{Ca}rlo radiative trans\textbf{fe}r code (MoCafe%
\footnote{http://kiseon.kasi.re.kr/MoCafe.html%
}, \citealt{Seon2009,SeonWitt2012,SeonWitt2013}) is used to perform
the dust scattering simulation of the UV scattered light. MoCafe is
a fully 3D radiative transfer model, which takes into account multiple
scattering. For the multiple dust scattering model, the Henyey\textendash Greenstein
scattering phase function, $\Phi\left(\theta\right)$ (\citealt{HenyeyGreenstein1941}),
is adopted: 

\begin{equation}
\Phi\left(\theta\right)=\frac{a}{4\pi}\frac{\left(1-g^{2}\right)}{\left(1+g^{2}-2g\cos\theta\right)^{\nicefrac{3}{2}}}.\label{Eq(1)}
\end{equation}

The albedo $a$ and the scattering asymmetric factor $g\equiv\left\langle \cos\theta\right\rangle $
define the scattering characteristics of the dust grains, where $\theta$
is the scattering angle of the incident photon. We also use the peeling-off
technique (\citealt{Yusef-Zadeh1984}) to improve the quality of the
resulting images. The code has been recently improved by adopting
a fast ray traversal algorithm (\citealt{Seon2015,AmanatidesWoo1987}).
Dust scattering is more efficient in UV light than in optical wavelengths.
Even a region of the $\lambda$ Ori nebula that is optically thin
in visible $\left(E(B-V)\approx0.1\right)$ produces an optical depth
in UV light of approximately 1, and thus the scattered UV radiation,
originating from the central O-type star $\lambda$ Ori, can be easily
detected in the nebula region (\citealt{Morgan1980}). 

The advantage of using scattered light to study the 3D structure of
clouds is that the dust scattering is strongly forward throwing at
UV wavelengths (e.g., \citealt{Witt1997,Draine2003}) and the scattered
intensity is very sensitive to the dust structure in front of the
light source. We conduct radiative transfer simulations for the shell
and the ring models; we compare the results with those from UV observation
in order to clearly demonstrate the presence of a shell structure.
We assume three types of cloud geometry, shown in Figure \ref{Fig2}.
The details of the models and the results are described in the following
subsections. 

\begin{figure}[t]
\begin{centering}
\includegraphics[width=7cm]{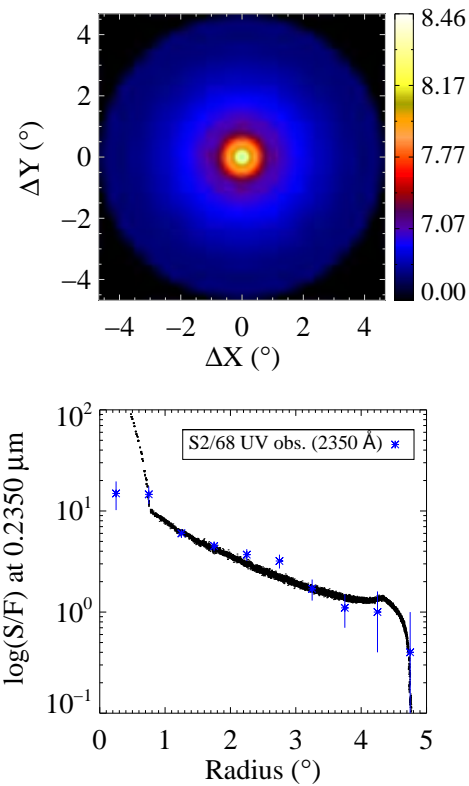}
\par\end{centering}

\protect\caption{\label{Fig3}Results of dust radiative transfer with the simple shell
model (Figure \ref{Fig2}(a)) with $a=0.37$ and $g=0.58$. The top
panel provides an image of UV scattered light (in units of erg cm$^{-2}$
s$^{-1}$ Å$^{-1}$ sr$^{-1}$) at 2350 Å, centered on $\lambda$
Ori. The color scale of the scattered light is given by asinh(F$_{i}$/median(F)),
where F$_{i}$ is the original flux at each pixel. The bottom panel
shows the ratios of surface brightness to incident stellar flux (log(S($\lambda$)/F$_{*}$($\lambda$)))
versus radius. Blue asterisks and black dots denote the observational
data and our results, respectively. }
\medskip{}
\end{figure}

\begin{table}
\protect\caption{\label{Table1} Optical properties of dust grains in $\lambda$ Ori
nebula}

\begin{centering}
\begin{tabular}{lcccc}
\hline 
\hline Properties & \multicolumn{4}{c}{Wavelength (Å)}\tabularnewline
\cline{2-5} 
 & 2740 & 2350 & 1950 & 1550\tabularnewline
\hline 
$a$ (shell$^{1}$) & $0.56{}_{-0.17}^{+0.14}$ & $0.37_{-0.04}^{+0.04}$ & $0.31_{-0.05}^{+0.03}$ & $0.38_{-0.03}^{+0.03}$\tabularnewline
$a$ (shell+ring$^{1}$) & $0.50_{-0.15}^{+0.13}$ & $0.29_{-0.03}^{+0.04}$ & $0.26_{-0.04}^{+0.03}$ & $0.31_{-0.03}^{+0.03}$\tabularnewline
$a$ (Morgan$^{2}$) & $0.52\pm0.16$ & $0.28\pm0.04$ & $0.24\pm0.06$ & $0.30\pm0.03$\tabularnewline
$g$ (shell$^{1}$) & $0.66_{-0.23}^{+0.15}$ & $0.58_{-0.12}^{+0.10}$ & $0.64_{-0.12}^{+0.10}$ & $0.63_{-0.07}^{+0.06}$\tabularnewline
$g$ (shel+ring$^{1}$) & $0.62_{-0.30}^{+0.17}$ & $0.61_{-0.12}^{+0.11}$ & $0.64_{-0.13}^{+0.11}$ & $0.65_{-0.06}^{+0.06}$\tabularnewline
$g$ (Morgan$^{2}$) & 0.50 & 0.50 & 0.50 & 0.50\tabularnewline
\hline 
\end{tabular}
\par\end{centering}

\textbf{Notes.} Albedo ($a$) and scattering asymmetry factor ($g$)
of the dust grains in $\lambda$ Ori nebula. References 1 and 2 refer
to the present models and the \citet{Morgan1980}'s shell model, respectively.
\end{table}

\subsection{Shell Structure Model }

The structure of the shell model and the luminosity of the central
source, located at a distance of 400 pc from the Earth (\citealt{Murdin1977}),
are defined following \citet{Morgan1980}. Morgan divided the nebula
structure into three concentric spherical shells in accordance with
the observed morphology: a central core cloud whose radius is 5 pc,
an outer shell that has a thickness of 3 pc (radius $\approx$ 30
to 33 pc), and a diffuse medium located between the central core and
the outer shell (Figure \ref{Fig2}(a)). To calculate the density
distribution of this simple model, the total extinction per unit length,
determined by means of star counts at the $B$ band, is set at $\approx$
0.086 mag/pc for the core, at $\approx$ 0.0028 mag/pc for the diffuse
medium, and at $\approx$ 0.05 mag/pc for the outer shell (\citealt{Coulson1978,Morgan1980}).
In our simulation, these $A_{B}$ values per unit length are converted
into values at four UV wavebands by using dust extinction cross-sections
(\citealt{WeingartnerDraine2001,Draine2003}) for Milky Way dust.
The results of the shell model at 2350 Å, using $10^{8}$ photons
for the central source, are shown in Figure \ref{Fig3}. Although
the adopted model structure is simple, having only three intervals
of density profile, the results accord fairly well with the observations
and with Morgan's results \citeyearpar{Morgan1980}.\textbf{ }Model
results at the other three UV bands show agreement within the observational
errors as well.

The main difference between the Morgan's results and ours for the
shell model is that we obtained slightly higher albedos, as shown
in Table \ref{Table1}. Morgan used the values from \citet{Allen1973},
in which the extinction cross-section is $\kappa$ = $10^{-9}$ cm$^{2}$,
independent of wavelength, and dust grain mass is $2\times10^{-13}$
g \citep[see also][]{Coulson1978}. These values give a relatively
higher dust density than that derived by the Milky Way dust model
from \citet{WeingartnerDraine2001}; consequently, our model requires
higher albedo to reproduce a level of scattered light close to that
found in the observations.

\begin{figure*}[t]
\begin{centering}
\includegraphics[scale=0.9]{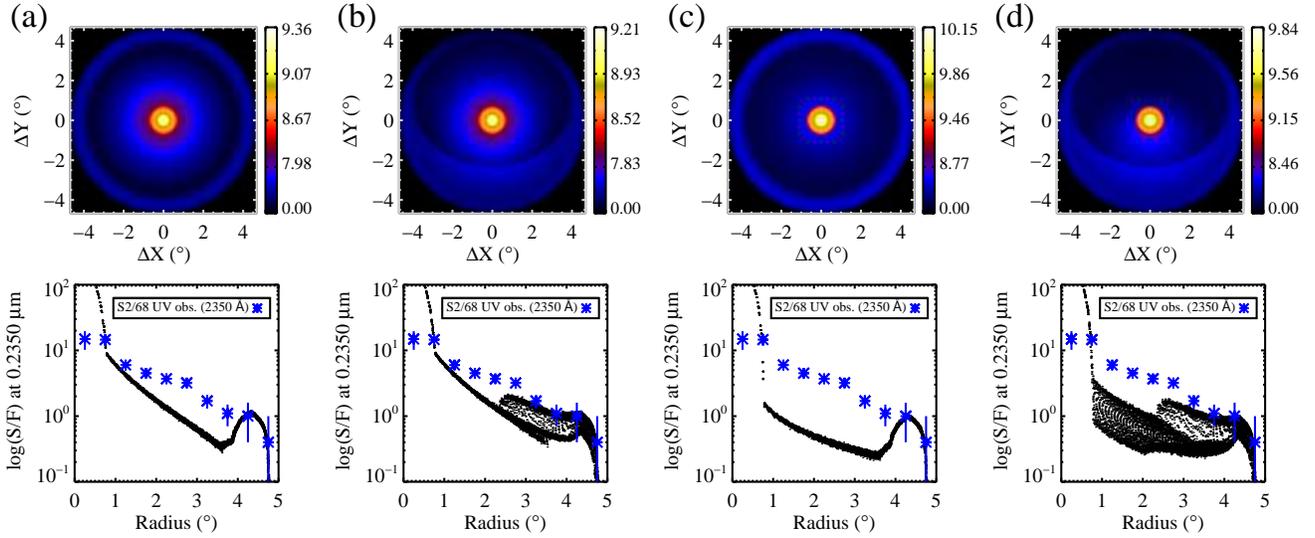}
\par\end{centering}

\protect\caption{\label{Fig4} Same as Figure \ref{Fig3}, but for the ring models
($a=0.37$ and $g=0.58$): (a) and (b) are for ring model 1, (c) and
(d) are for ring model 2. Figures (b) and (d) are the results when
the inclination of 30\textdegree{} is applied. The structure of each
ring model is shown in Figure \ref{Fig2}. }
\medskip{}
\end{figure*}

\subsection{Ring Structure Model}

We define two additional ring models as shown in Figures \ref{Fig3}(b)
and (c). In these models, we introduced an opening angle $\beta$
of the ring structure. Both models are circularly symmetric about
the $x$-axis and have a ring structure with opening angle $\beta$
at the edge of the nebula. Ring model 1 (Figure \ref{Fig2}(b)) has
the same diffuse medium as the shell model (Figure \ref{Fig2}(a)),
while ring model 2 (Figure \ref{Fig2}(c)) has that of ring type with
the same opening angle $\beta$ as that of the outer ring structure.
The diffuse medium is colored a brighter grey in each figure. Note
that the shell model has $\beta$ of 90\textdegree . The same total
optical depth is applied to both the shell and ring models and thus
the dust densities of the diffuse medium and core in the ring models
are slightly higher than those of the shell model.

The ring model results for $\beta$ = 30\textdegree{} at 2350 Å are
shown in Figure \ref{Fig4}. Figures \ref{Fig4}(a) and (b) show results
of ring model 1; Figures \ref{Fig4}(c) and (d) show those of ring
model 2. As can be clearly seen in the figures, compared to the observational
data denoted by blue asterisks, the ring model results (black dots)
underpredict the intensity in a certain radial range between the core
and the outer boundary (1\textdegree{} < $r$ < 4\textdegree ). Ring
model 2 results in a much weaker intensity than that of ring model
1. The same phenomenon was found in the ring models with different
values of $\beta$, such as 10\textdegree , 20\textdegree , 40\textdegree ,
and 50\textdegree . This implies that the diffuse medium alone, without
the outer shell structure, is not enough to explain the observed UV
scattered light, and that a geometrically thin but dense shell of
dust with a radius of about 30 pc is required. This is mainly due
to the fact that UV photons are strongly forward scattered and thus
the scattered light in the radial range of $r$ < 4\textdegree{} predominantly
originates from a dust cloud located in front of the light source. 

\citet{Lang2000} reported on the inclination of the ring cloud with
respect to the plane of the sky, suggesting an upper limit of $\approx$
30\textdegree . The inclination of the ring cloud was incorporated
by rotating the observer location anti-clockwise about the $y$-axis
in Figure \ref{Fig2}, equivalent to a clockwise rotation of the cloud
system. Simulation results, assuming this inclination angle of about
30\textdegree , are shown in Figures \ref{Fig4}(b) and (d). The southern
part of the inclined ring, which is nearer to the observer, covers
about half of the southern region of the nebula and consequently plays
the role of dense shell, as one would expect in the shell model, producing
UV scattered light up to the level seen in the observation. In the
figures, the intensity appears to be split into two branches: one
corresponding to the northern part, with lower intensity, and the
other corresponding to the southern part, with higher intensity. The
lower intensity in the northern part is due to the fact that dust
on the far side of the ring scatters fewer photons to the observer.
Moreover, the inclined ring models still show a significant deficit
in scattered light, between 1\textdegree{} to 2\textdegree .5, where
the inclined ring cannot cover and scatter photons from the central
source. In conclusion, both the ring and the inclined ring models
fail to fit the UV observation data, while the shell model succeeds.
Accordingly, these model results strongly suggest that it is necessary
to adopt a shell structure in order to explain the UV scattering data.

\section{SHELL WITH RING STRUCTURE}

\begin{figure}
\begin{centering}
\includegraphics[width=5.5cm]{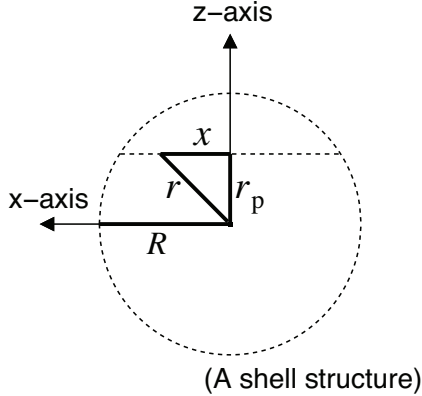}
\par\end{centering}

\protect\caption{\label{Fig5} Schematic diagram of spherically symmetric density profile.
Column density is a function of $r_{p}$; density, the result of the
inverse Abel transform, depends on $r$. The density becomes zero
outside of $R$, which is the maximum radius of the sphere. The coordinates
system is same as that shown in Figure \ref{Fig2}, and it is simply
assumed that each line of sight is parallel to the $x$-axis, so that
the projected radius, $r_{p}$, is defined on the $y$-$z$ plane. }
\medskip{}
\end{figure}

According to the Monte Carlo radiative transfer models in the previous
section, it is now clear that there should exist a dust shell cloud
around $\lambda$ Ori. In this case, the inverse Abel transform (\citealt{Binney2008})
can be used to estimate the 3D radial density profile of the dust
cloud using the spherical symmetry of the shell cloud. The Abel integral
equation relates a 3D density profile to its 2D column density for
a spherically symmetric structure: 

\begin{equation}
\sigma\left(r_{p}\right)=2\int_{r_{p}}^{R}\frac{\rho\left(r\right)r}{\sqrt{r^{2}-r_{p}^{2}}}dr,\label{Eq(2)}
\end{equation}

where $\sigma\left(r_{p}\right)$ is the observed column density (hereafter
denoted as column density or $\sigma$), $\rho\left(r\right)$ is
the radial volumetric density profile of a spherical cloud (hereafter
density or $\rho$), $r_{p}$ is the projected radius, $r$ is the
radial distance from the center, and $R$ is the outer radius of the
shell, as shown in Figure \ref{Fig5}. Assuming that $\rho\left(r\right)$
is continuously differentiable, the inversion of the Abel transform
is given by 

\begin{equation}
\rho\left(r\right)=-\frac{1}{\pi}\int_{r}^{R}\frac{d\sigma\left(r_{p}\right)}{dr_{p}}\frac{dr_{p}}{\sqrt{r_{p}^{2}-r^{2}}}.\label{Eq(3)}
\end{equation}

\begin{figure}
\begin{centering}
\includegraphics[width=8cm]{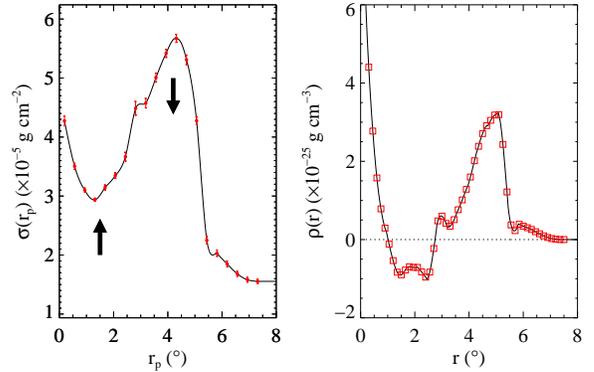}
\par\end{centering}

\protect\caption{\label{Fig6} The left panel shows the azimuthally averaged dust column
density versus projected radial distance from $\lambda$ Ori, with
standard errors. The column density when the projected radius is over
5\textdegree .2 is derived from the \textquotedblleft $bg$\textquotedblright{}
areas in the $E(B-V)$ map (Figure \ref{Fig1}). The right panel shows
the mass density of the $\lambda$ Ori nebula, obtained by calculating
the inverse Abel transform while using the assumption that the entire
nebula system is a shell. It should be noted that negative values
are seen in the radial density profile at $r\approx$ 2\textdegree . }

\medskip{}
\end{figure}

Figure \ref{Fig6} shows the azimuthally averaged column density derived
from the $E(B-V)$ map, and the radial density profile calculated
by applying the inverse Abel transform to the column density $\sigma\left(r_{p}\right)$.
However, we obtain negative density values in a certain radial range,
which is not physically possible. The absolute value is too great
to be numerical error ($\approx$ 30\% of the peak density at $r_{p}\approx$
5\textdegree ). Generally, the inverse Abel transform can produce
negative values in areas around the center if the input column density
is insufficient at the central region or excessive at the outer region.
Therefore, there can be two mathematical ways to solve this problem:
one is to enhance the input column density at $r_{p}\approx$ 1\textdegree $-$2\textdegree ;
the other is to reduce the input column density at $r_{p}\approx$
4\textdegree $-$5\textdegree{} (marked by arrows in the left panel
of Figure \ref{Fig6}). Both processes will make the negative values
rise; however, it is practically impossible to ``increase'' the
column density as we cannot add new materials to the present cloud
system. Accordingly, the only feasible way is to reduce the peak column
density. This reduction is physically possible when there is a certain
amount of dust along the outer part, which is not spherically symmetric
but coexistent with the shell structure. In other words, the negative
density problem can be resolved by assuming an additional non-spherical
dust cloud component that overlaps with the spherical shell. The non-spherical
cloud can be regarded as a toroidal ring-like dust cloud; then, the
ring component can be subtracted from the column density profile. 

\citet{Nielbock2012} modeled the column density of the dark globule
Barnard 68 using the equation 

\begin{equation}
\sigma\left(r_{m}\right)=\frac{\sigma_{0}}{1+\left(r_{m}/r_{m_{0}}\right)^{\alpha}},\label{Eq(4)}
\end{equation}

where $r_{m}$ is the radial distance from the center of the globule,
$\sigma_{0}$ is the central column density, $r_{m_{0}}$ is a certain
radius that determines the range of an inner flat density core at
$r_{m}$ < $r_{m_{0}}$, and $\alpha$ controls the power-law steepness
of the profile. We describe the value of $\sigma\left(r_{m}\right)$
of the ring structure around $\lambda$ Ori, assuming this equation
can plausibly fit not only the surface density profile of a single
globule but an averaged profile of overlapped clumps along the ring.
The profile was originally proposed to model the spherically symmetric
globule, but we will apply it to this case for convenience. The radial
distance $r_{m}$ in the equation can now be applied as the radius
of a cross-section of the ring (minor radius), while the projected
radial distance $r_{p}$ can be used as the distance from the center
of the torus tube (major radius), which is located at a distance 4\textdegree .2
or 4\textdegree .6 from the nebula center. For better fitting, Equation
(\ref{Eq(4)}) is separately applied to Region 1 and 2 (Figure \ref{Fig1}),
since these two regions have different features of clumps in terms
of apparent size and brightness. In addition to these rings, we especially
consider constructing an additional ring cloud that includes a cloud
complex known as B30, which has very bright emissions in both CO and
IR (\citealt{Lang2000}). It should be noted that the radial distance
to the center of this complex is about 1\textdegree{} shorter than
that of the other clumps of the ring in Region 1. This consequently
creates a bump in the column density profile at $r_{p}\approx$ 2\textdegree .75,
which needs to be subtracted in order not to produce a greater negative
density profile. It is likely that the bump in the UV observation
data at the same radius would be due to this complex. Here in the
present study, an arc-shaped ring cloud whose angular size is about
90\textdegree{} is newly defined. This partial ring structure, which
we term the ``B30 arc complex,'' includes the B30 and other clumps
that are located at about 2\textdegree .75 from $\lambda$ Ori (indicated
by the black dashed ellipses in Figure \ref{Fig1}). 

\begin{figure*}[t]
\begin{centering}
\includegraphics[scale=0.9]{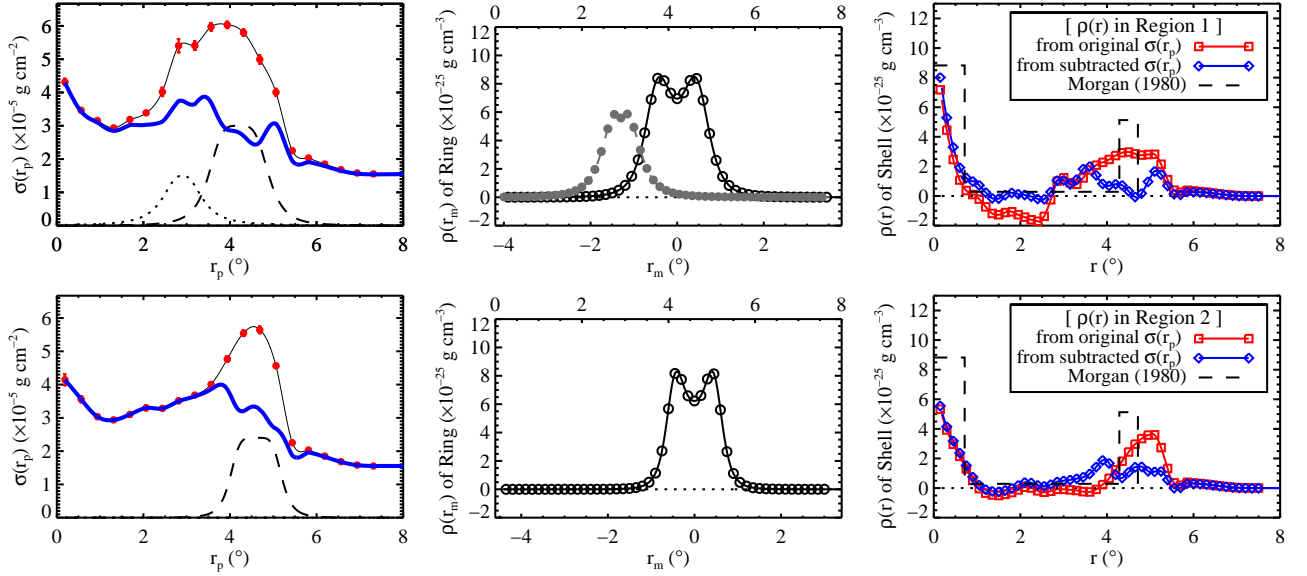}
\par\end{centering}

\protect\caption{\label{Fig7} The ring and shell density profiles. The top and bottom
panels are for Region 1 and Region 2, respectively (see Figure \ref{Fig1}).
The left panels show azimuthally averaged profiles of the dust column
density (red dots and their interpolated solid lines, the same as
those shown in the left panel of Figure \ref{Fig6}), the ring (dashed
lines) and the B30 arc complex structure (dotted lines), and the ring-subtracted
structure (thick blue lines). The middle panels show density profiles
of the ring (open circles) and B30 arc complex (filled circles), plotted
against the minor radius of the ring, $r_{m}$, which is centered
on a cross-section of the ring cloud. The upper axes of the middle
panels represent the relative position of the ring\textquoteright s
major radius ($r_{p}$). The density profiles of the spherically symmetric
dust clouds are displayed in the right panels. The red squares are
derived from the original column density; the blue diamonds are derived
from the ring-subtracted column density; the dashed lines indicate
the simple shell model structure (\citealt{Morgan1980}), seen in
Figure \ref{Fig2}(a), which produces scattered light, as shown in
Figure \ref{Fig3}. Note that the radial coordinates of the ring clouds
in the left panels are not the radial distances from the center of
the nebula; rather, they are projected distances of the torus tubes
on the $y$-$z$ plane, as shown in Figure \ref{Fig2}. The coordinates
in the middle panels are measured from the center of the torus tubes
of the ring cloud. }

\medskip{}
\end{figure*}

\begin{figure}
\begin{centering}
\includegraphics[width=7cm]{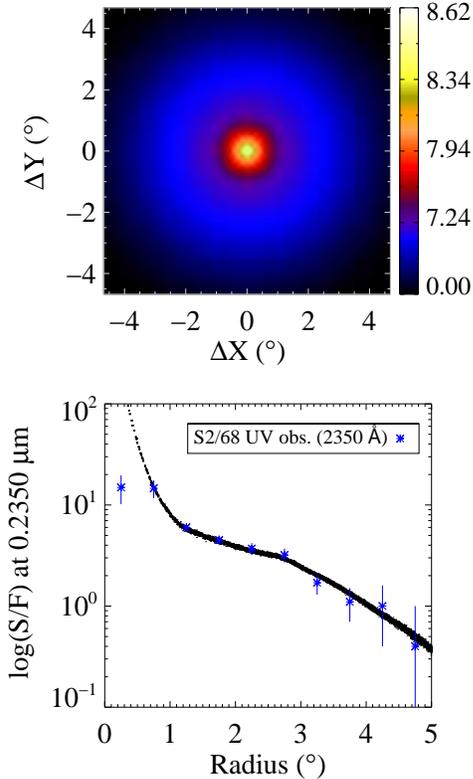}
\par\end{centering}

\protect\caption{\label{Fig8}Same as Figure \ref{Fig3}, but the best-fit result using
both the ring and the shell density profiles defined in Figure \ref{Fig7}
and Table \ref{Table1} ($a=0.29$ and $g=0.61$ at 2350 Å). }
\medskip{}
\end{figure}

Using the column density of the ring structure and the B30 arc complex
(dashed and dotted lines, respectively, in the left panels of Figure
\ref{Fig7}), we obtained the column density (blue solid lines in
the same panels) and $\rho\left(r\right)$ of the ``pure'', spherically
symmetric structure (blue diamonds in right panels of Figure \ref{Fig7})
using the inverse Abel transform. The negative shell density has been
remarkably reduced and is now close to zero due to the ring-subtracted
column density. There are no perfect, spherically symmetric clouds;
such imperfection would cause the errors in the $\rho\left(r\right)$
profile. In practice, however, it may be possible to find a density
profile that is reasonable within a certain limit. Therefore, we define
a limit of tolerance, 5 percent of the maximum density, and permit
some negative values of $\rho\left(r\right)$ within this tolerance
limit. These small negative values were set to zero density in the
dust radiative transfer code. We also derive $\rho\left(r\right)$
of the ring cloud (the middle panels of Figure \ref{Fig7}) using
the truncated off-center Gaussian function (\citealt{Ciotti2000}).
The parameters of Equation (\ref{Eq(4)}), describing the ring structures,
are tabulated in Table \ref{Table2}. The simulation result for the
scattered light obtained by using the shell and ring densities, averaged
over the azimuth angles, is shown in Figure \ref{Fig8}. This result
shows much better agreement with the UV observation than was found
in previous studies. Only the central region, where the azimuthally
averaged column density may not be reliable due to the small quantity
of the pixel data, still shows an inconsistency, as was found in previous
works (\citealt{Morgan1980}). The best-fit optical properties of
the dust grains in the $\lambda$ Ori nebula are listed in Table \ref{Table1}.
Interestingly, the albedos obtained from our final (shell+ring) model
are consistent with those of the shell model of \citet{Morgan1980}
whereas our shell model in the previous section yielded higher values.
One thing we should note is that the final radial density at the outer
boundary of the spherically symmetric dust cloud is rather low compared
to that of the ring cloud. However, as noted in the discussion, the
total dust mass included in the outer shell region is higher than
that in the ring dust cloud.

\begin{table}
\protect\caption{\label{Table2} Parameters of ring structures and B30 arc complex }

\begin{centering}
\begin{tabular}{llcccc}
\hline 
\hline Ring structures &  & Major radius (\textdegree ) & $\sigma_{0}$ (g cm$^{-2}$) & $r_{p_{0}}$ (\textdegree ) & $\alpha$\tabularnewline
\hline 
Ring in Region 1 &  & 4.2 & $3.0\times10^{-5}$ & 0.7 & 4.0\tabularnewline
Ring in Region 2 &  & 4.6 & $2.4\times10^{-5}$ & 0.5 & 5.0\tabularnewline
B30 arc complex &  & 2.9 & $1.5\times10^{-5}$ & 0.5 & 2.5\tabularnewline
\hline 
\end{tabular}
\par\end{centering}

\textbf{Notes.} Input parameters describing the azimuthally averaged
column density of the ring and the B30 ring complex. $\sigma_{0}$,
$r_{p_{0}}$, and $\alpha$ are defined in Equation (\ref{Eq(4)}).
\end{table}

\section{DISCUSSION}

\subsection{$\lambda$ Ori Nebula as a Ring with a Shell Structure}

The present study was designed to study the 3D structure of the $\lambda$
Ori nebula; more specifically, to examine the existence of a shell
and/or a ring structure in the nebula. The previous radiative transfer
models (i.e., \citealt{Morgan1980}), based on star counts, assumed
a simple density structure, divided into three concentric shells.
We, however, directly used a recently observed dust reddening map
(\citealt{Schlafly2014a}), along with the inverse Abel transform
technique, which enables us to envision a spherically symmetric radial
density profile of the shell cloud. Negative density profiles are
found when no ring structure is considered, indicating that the dust
clouds are composed of two structures, a spherical shell and a toroidal
ring. The scattered light, calculated using the Monte Carlo radiative
transfer model, for both shell and ring structures, nicely reproduced
the UV observations, including even the bump profile seen at 2\textdegree .75
(Figure \ref{Fig8}), which has not been reproduced in any previous
analyses. These results, in turn, strongly suggest that the $\lambda$
Ori nebula consists of both ring- and shell-like dust clouds. 

The optical properties of the dust associated with the $\lambda$
Ori nebula at four UV bands were derived (Table \ref{Table1}). As
discussed in Section 3.1, compared to the data used by \citeauthor{Morgan1980},
we have used more refined dust extinction properties from relatively
more recent works in the literature. We have obtained similar albedo
but stronger forward scattering in all four UV bands than were found
by \citet{Morgan1980}. While Morgan\textquoteright s \citeyearpar{Morgan1980}
simulation, with an asymmetry factor of $g=0.50$, could not reproduce
the steep slope (see Figure 3 in \citealt{Morgan1980}), our simulation,
with a stronger forward scattering ($g=0.61$) accords well with the
radial profile shown in Figure \ref{Fig8}. 

\citet{SeonWitt2013} showed that dust scattering in turbulent ISM
can yield substantial scatter in the correlation plots between the
scattered flux and the optical depth. This effect is shown to be more
severe as the sonic Mach number increases (turbulent ISM) and the
scattering asymmetry factor of the dust grains decreases (isotropic
scattering ISM). This may alter the results of the radiative transfer
models using a relatively smoothly varying density. The effect is
essential in near-infrared wavelengths, in which dust scattering is
very close to isotropic, given normal submicron-size dust grains in
the diffuse ISM (\citealt{Draine2003}). As listed in Table \ref{Table1},
the scattering asymmetry factors of the $\lambda$ Ori dust cloud
are between 0.61 and 0.65, meaning strong forward scattering in the
UV wavelengths. Therefore, the ``weak correlation effect'' is negligible
in our scattering model, though the expanding $\lambda$ Ori system
may be turbulent due to a shock wave.

\citet{Mathis2002} demonstrated that observations of a hierarchically
clumped reflection nebula can be fitted by a wide range of albedos
when a smooth density structure is assumed. In our paper, we compared
a radially averaged 1D intensity map with the radiative transfer models
adopting a smooth density structure. Therefore, further investigation
with spatially resolved 2D images may be needed to better constrain
the optical properties.

The distance from the Sun to the target is an essential parameter
used to estimate the mass and size of the nebula. We have adopted
the distance of $D=400\pm40$ pc, proposed by \citet{Murdin1977}.
Since their study, several measurements for the distance to $\lambda$
Ori have been carried out. The distance derived from the parallax
of the Hipparcos catalog (ESA 1997) is $D=380\pm30$ pc; the distance
suggested by \citet{DolanMathieu2001} was $D=450\pm50$ pc, determined
in a photometric study. More recently, the distance to the $\lambda$
Ori nebula was estimated to be $D=420\pm42$ pc using a large catalog
of distances to the Galactic molecular clouds (\citealt{Schlafly2014b}).
The distance $D=400$ pc that we assumed is within these uncertainty
ranges. In fact, the distance to $\lambda$ Ori is not critical for
the determination of the three-dimensional structure of this nebula.
It will become possible to derive accurate distance and relevant physical
properties through data from the Gaia mission, launched at the end
of 2013.

\subsection{The Parent Cloud and Mass Estimation}

It is generally noted that interstellar gas and dust in bubble structures,
including those of the $\lambda$ Ori nebula, are swept up by a shock
wave. For the case of the $\lambda$ Ori system, a stellar wind (\citealt{Maddalena1987})
or a supernova (\citealt{CunhaSmith1996,DolanMathieu2002}), although
which one actually powered this shock wave is still unclear, presumably
has swept up the central materials and resulted in the current morphology.
Therefore, the current morphology of the nebula might be able to give
a clue to the original structure of the parent cloud. \citet{Coulson1978}
and \citet{Morgan1980} proposed that the current morphology is a
simple dust shell, suggesting that its original structure was spherical.
Meanwhile, \citet{Maddalena1987} suggested an oblate molecular cloud
as the birthplace of $\lambda$ Ori. In order to explain the CO observations,
which show a clear ring feature, they assumed a preexisting natal
cloud, which was not initially spherical but flat. The expanding \ion{H}{2}
region could explain the ring of CO clumps torn from the parent cloud.
With photometry of over 320,000 stars, \citet{DolanMathieu2002} studied
the spatial distribution of the pre-main-sequence stars around $\lambda$
Ori; they concluded that the progenitor cloud may have been elongated.
Hence, our simulation results, which clearly reveal the presence of
a ring structure, support the elongated or the oblate progenitor model
rather than the spherical one. 

The dust mass of the $\lambda$ Ori nebula can be derived by using
the volumes of the shell and the ring structures, and the density
profiles, shown in Figure \ref{Fig7}. We numerically integrated the
density profile and obtained the masses of the dust clouds: $\approx180$
$M_{\odot}$ for the spherical dust cloud, $\approx130$ $M_{\odot}$
for the dust rings including the B30 arc complex, and $\approx310$
$M_{\odot}$ in total. If we use a gas-to-dust ratio of 100, then
the total mass of the neutral gas in the shell is $\approx1.8\times10^{4}$
$M_{\odot}$ and that in the ring is $\approx1.3\times10^{4}$ $M_{\odot}$.
This ring mass is comparable to previous mass estimates: a dust ring
mass of $\approx170$ $M_{\odot}$ (\citealt{Zhang1989}), and a molecular
ring mass of $\approx2.8\times10^{4}$ $M_{\odot}$ (\citealt{Maddalena1987})
and that of $\approx1.4\times10^{4}$ $M_{\odot}$ (\citealt{Lang2000}).
The mass estimated in this study also points out that the shell structure
is significant in terms of mass compared to the ring, even though
the radial density of the shell at the boundary appears to be lower
than that of the ring, as can be seen in Figure \ref{Fig7}. 

If we adopt a dust reddening $E(B-V)$ of $\approx0.1$ toward $\lambda$
Ori and the extinction cross-section from \citet{WeingartnerDraine2001},
the density of the diffuse background medium near the $\lambda$ Ori
system is estimated to be $\approx8.83\times10^{-27}$ g cm$^{-3}$.
If there was no massive progenitor cloud around $\lambda$ Ori, but
diffuse dust medium with this density, and the \ion{H}{2} region
had swept up the diffuse medium to the shell's inner radius of 30
pc, then the mass of the swept dust shell would be approximately 15
$M_{\odot}$. Such a mass is more than an order of magnitude smaller
than the dust shell mass estimated in this study. Therefore, our results
support the hypothesis of a parent cloud around $\lambda$ Ori.

\subsection{A New Methodological Approach to the Study of the Morphology of Bubbles }

With CO $\left(J=1\rightarrow0\right)$ observations, \citet{BeaumontWilliams2010}
studied 43 bubbles and found a ring of cold gas, not a shell, around
them, suggesting that the parent molecular clouds had been flattened.
A comparison was made between the radial intensity profiles for the
shell model and for the CO data; comparison results strikingly show
the rarity of emission toward the center of the bubbles. They found
weak correlation between the shell model and the CO data and no convincing
evidence of front or back faces of the expanding shells. As a result,
they concluded that shock waves from massive stars tend to create
ring clouds instead of spherical shells. However, it should be noted
that this ring morphology indicates a structure of ``molecular gas
clouds,'' not of ``entire cloud systems'' including dust and neutral
gas. According to our results, the $\lambda$ Ori system has not only
dust ring but also dust shell structures. This interpretation stresses
that one should not overlook investigating dust morphology in order
to determine the structure of an ``entire cloud system,'' and that
our approach can be a useful one to understand the 3D structure of
bubbles. 

It is normally assumed that the morphology of well-defined ring-like
bubbles can be described by one of two structures: a ring or a shell
(or both). A relatively dilute shell cloud surrounding a nebula may
not be detectable with CO and/or \ion{H}{1} observations. It should
be noted that two previous \ion{H}{1} studies of the $\lambda$ Ori
nebula reported controversial interpretations: \citet{Wade1957} proposed
a nearly spherical \ion{H}{1} shell, while \citet{ZhangGreen1991}
concluded that the \ion{H}{1} shell that corresponds to the ring
is far from complete. In this controversial case, the 3D radiative
transfer model for scattered light can provide a solution because
dust scattering is forward-directed in the optical and UV wavelengths,
and thus this method is sensitive to the existence of shell dust in
front of the central source. However, this method is not good at probing
the ring dust cloud, as demonstrated in the present study. It should
be noted that even a simple shell model without ring structures gave
rather good agreement with their observational data, as can be seen
in Figure \ref{Fig3}. The existence of the ring dust cloud could
be found by negative density values estimated using the inverse Abel
transform. In the end, we obtained an integrated view on the structure
of the $\lambda$ Ori nebula: a cloud with both shell and ring structures.

Our model results and the method are not perfect for several reasons.
First, the inverse Abel transform deals with an ill-posed problem.
In fact, clumpy ring clouds cannot be represented by a simple ring
model (Equation (\ref{Eq(4)})). The small bumps remaining after subtracting
the ring column density can distort the output density profile. However,
this effect can be reduced by using a priori information and a smooth
input function (\citealt{Craig1979}). To increase stability we have
divided the nebula into a proper number of azimuthal intervals and
used the averaged column densities in each of the sub-regions. In
addition, a priori information on spherical and circular symmetries
and, above all, the constraint of positive definiteness, were used.
Although the detailed shape of the output density profile was changed
in attempting to determine the various ring dust clouds profiles,
the presence of both shell and ring structures could be clearly seen.
Second, the UV observation data we used were not 2D images, but a
1D radial profile. If, therefore, the UV image for the target field
can be provided in the future, a more accurate density structure of
the $\lambda$ Ori system will be derived. 

The new approach described in this paper, using both the radiative
transfer model and the inverse Abel transform, has allowed us to detect
the presence of both dust shell and ring clouds. Given a full map
of dust reddening and UV data, it is likely that this approach can
be applied to other relatively symmetrical bubbles centered on ionizing
sources. There have been many studies of bubble-like structures over
various scales (\citealt{Castor1975,Heiles1979,ReynoldsOgden1979,MacLowMcCray1988,Kendrew2012,Simpson2012}),
suggesting rounded bubbles are common morphology in the ISM. Accordingly,
we may apply our method to many bubbles to examine whether their dust
structures are of shell or ring type, or both, if they have massive
or hot young stars and are not overlapped with unwanted massive clouds
along the line of sight.

\section{SUMMRARY}

We successfully applied a new methodological approach, using a 3D
Monte Carlo radiative transfer model and the inverse Abel transform,
to study the structure of the $\lambda$ Ori nebula, which structure
has been controversial for decades. Our results show that a ring cloud
coexists with a dust shell structure, suggesting there was a flat
or elongated progenitor cloud. Although the morphology of the neutral
and molecular hydrogen cloud in the target field could not be revealed
through the present study, we suggest that our technique may be applicable
to studies of bubble structure. Together with the results of previous
studies based on CO and IR measurements, our technique may enable
us to develop an integrated interpretation of the morphology of the
$\lambda$ Ori system.

\end{document}